\documentclass[11pt]{article}
\usepackage{amssymb,graphicx,epsfig,cite,hyperref} 
 \evensidemargin=\oddsidemargin 
\def\baselinestretch{1.2}

\def\lsim{\ \raisebox{-.4ex}{\rlap{$\sim$}} \raisebox{.4ex}{$<$} \ } 
\def\gsim{\ \raisebox{-.4ex}{\rlap{$\sim$}} \raisebox{.4ex}{$>$} \ } 
\begin{document}
\begin{titlepage}
\begin{flushright}
TIFR/TH/12-15
\end{flushright}

\begin{center}
{\LARGE\bf
How Constrained is the cMSSM?} \\[5mm]
\bigskip
{\large\sf Diptimoy Ghosh} $^{a,1}$, 
{\large\sf Monoranjan Guchait} $^{b,2}$, 
{\large\sf Sreerup Raychaudhuri} $^{a,3}$ 
and
{\large\sf Dipan Sengupta} $^{b,4}$ 
\\ [4mm]
\bigskip

{\noindent $^{a)}$ 
Department of Theoretical Physics, 
Tata Institute of Fundamental Research,  \\
\hspace*{0.1in} 1, Homi Bhabha Road, Mumbai 400 005, India. }

\medskip

{\noindent $^{b)}$ 
Department of High Energy Physics, 
Tata Institute of Fundamental Research,  \\
\hspace*{0.1in} 1, Homi Bhabha Road, Mumbai 400 005, India. }
\end{center}

\vskip 50pt
\begin{center}
{\large\bf ABSTRACT}
\end{center}

\begin{quotation} 
\noindent We study the allowed parameter space of the constrained 
minimal supersymmetric Standard Model (cMSSM) in the light of direct 
searches, constraints from $B$-physics (including the recent measurement 
of the branching ratio for $B_s \to \mu^+\mu^-$) and the dark matter 
relic density. For low or moderate values of $\tan\beta$, the strongest 
constraints are those imposed by direct searches, and therefore, large 
areas of the parameter space are still allowed. In the large $\tan 
\beta$ limit, however, the $B$-physics constraints are more restrictive, 
effectively forcing the squark and gluino masses to lie close to or 
above a TeV. A light Higgs boson could dramatically change the allowed 
parameter space, but we need to know its mass precisely for this to be 
effective. We emphasize that it is still too early to write off the 
cMSSM, even in the large $\tan\beta$ limit. Finally we explore 
strategies to extend the LHC search for cMSSM signals beyond the present 
reach of the ATLAS and CMS Collaborations.
\end{quotation}

\vskip 60pt 

\begin{center} 
PACS numbers: {\tt 12.60.Jv, 14.40.Nd, 13.85.Rm}
\end{center}

\vfill

\begin{center}
\today
\end{center}

\hrule
\vskip -10pt
\footnotesize \noindent
\centerline{
$^1$diptimoyghosh@theory.tifr.res.in \qquad
$^2$guchait@tifr.res.in \qquad
$^3$sreerup@theory.tifr.res.in \qquad
$^4$dipan@tifr.res.in} 
\normalsize
\end{titlepage}
\newpage
\setcounter{page}{1}

\section{Introduction}

\vskip -10pt

The early results pouring in from the CERN Large Hadron Collider (LHC) 
and the remarkable speed and efficiency with which they are being 
analysed for their physics potential represents a major triumph for the 
smooth operation of the machine and for the meticulous planning that has 
gone into every facet of the programme. Even though the LHC has utilised 
but a small fraction of its energy and luminosity potential, LHC data 
have already overtaken the LEP and Tevatron results in many search 
channels. More significantly, apart from some excitement about a 
possible 125~GeV Higgs boson \cite{cmshiggs,atlashiggs}, it is clear 
from the LHC results presented during the past year or so that there is 
no new physics `around the corner', and we may have to settle for a long 
and suspenseful wait before the experiment achieves its first 
breakthrough in this regard. However, this is no reason for despair, for 
the LHC still has a long way to go before any serious verdict can be 
pronounced on the different new physics options currently under 
consideration.

In this article, we discuss the impact of the latest experimental data 
on the 'constrained' minimal supersymmetric Standard Model (cMSSM) 
\cite{cmssm}. Some time ago, the cMSSM was often dubbed as the `standard 
model' of physics beyond the Standard Model (SM), but perhaps because of 
the string of negative results obtained so far, there has lately arisen 
a tendency to disparage the cMSSM as a model which makes too many 
arbitrary assumptions and hence is --- not-surprisingly --- on the verge 
of getting ruled out by the LHC data \cite{atlaslimit,cmslimit}. Such 
views, are however, less than fair to the cMSSM, for, in many ways, the 
cMSSM may be regarded as the most simple and economical model of 
supersymmetry, and the one which would most obviously suggest itself in 
the absence of contradictory experimental evidence. In fact, if one 
thinks about it, we should rather regard the multifarious alternatives 
to the cMSSM which appear in the literature as the ones where extra 
assumptions are introduced. Moreover, the mere fact that the cMSSM 
parameter space is getting reduced by experimental searches should not 
be regarded as a setback for the model, for, after all, Nature 
corresponds to but a single point in the parameter space. The example of 
the top quark (and maybe the Higgs boson) serves to clearly illustrate 
this kind of shrinkage of the parameter space to the actual value.

The purpose of this article is not, however, to pontificate in defence 
of the cMSSM, but rather to investigate the status of different 
experimental constraints on this model. Among others, we take up the 
recent measurement of the process $B_s \to \mu^+ \mu^-$ by the LHCb 
Collaboration \cite{Aaij:2012ac} and study its impact on the cMSSM 
parameter space in conjunction with other low energy constraints. Since 
the precision of this particular measurement has increased considerably, 
one would expect it to rule out a wide swath of the parameter space. We 
quantify this expectation and find, that while the last statement is 
certainly true at large values of $\tan\beta$, the constraint weakens 
and disappears as $\tan\beta$ is lowered. We shall demonstrate that even 
in the large $\tan\beta$ region, the cMSSM is still a possible 
explanation of not just the hierarchy problem, but also of the dark 
matter problem. Finally we shall use a novel strategy using event-shape 
variables, developed by two of the authors \cite{dipan}, to extend the 
experimental search for the cMSSM to a region of parameter space which 
is allowed by all the present constraints, but inaccessible to 
conventional searches.

The plan of this article is as follows. In the next section we briefly 
discuss the cMSSM and then go on to discuss how this is constrained. 
This is followed by a section in which we map the allowed parameter 
space by including all the relevant constraints, and thus, focus on the 
viability or otherwise of the model. Some comments on the impact of a 
125~GeV Higgs boson are also included in this section. We then go on to 
describe our novel search strategy, which enables us to extend the 
search for cMSSM particles into new regions of the parameter space. A 
critical summary of our work forms the concluding section.

\vspace*{-0.2in}

\section{The Model and its Constraints}

\vskip -10pt
The cMSSM, whose roots go back to the 1980's, now forms the material of 
textbooks\cite{cmssm,martin}, and hence, our description of the model 
will touch only upon its salient features. The particle content of this 
model may be summarised as a two-Higgs doublet extension of the SM plus 
superpartners, for all the particles therein. These will have 
interactions determined by the SM interactions, except for the soft 
SUSY-breaking sector, which consists of a set of terms included in an ad 
hoc manner, with the only constraint that they should not lead to 
quadratic divergences in the mass of the lightest Higgs boson.

It is postulated that there exists a hidden sector of fields which do 
not interact with the known SM fields through the known SM interactions. 
In this hidden sector, SUSY is broken spontaneously, at the cost of 
having massless goldstinos, which are, however, invisible, being in the 
hidden sector\footnote{In supergravity models, this goldstino is 
absorbed into the massive gravitino.}. However, the two sectors can 
interact through gravity, which is universal, and thus if we integrate 
out the hidden sector fields from the gravity-mediated interactions 
terms we are left with effective interactions which are just the soft 
SUSY-breaking terms. This will take place at some high energy scale 
below the Planck scale. At this scale, all the soft SUSY-breaking terms 
are generated purely by gravitational interactions, which are blind to 
all the SM quantum numbers, such as colour, flavour, etc. and sensitive 
only to space-time quantum numbers such as energy, momentum and spin. 
Accordingly, at this scale, we must assume that there is a common scalar 
mass $m_0$, a common fermion mass $m_{1/2}$ and a common trilinear 
coupling $A_0$ in addition to the Higgsino-mixing parameter $\mu$. Any 
splitting among these (known as {\sl non-universality}) would require 
the hidden sector to be sensitive to the global SM quantum numbers (e.g. 
flavour) which distinguish the different visible fields. We do not make 
this assumption in the cMSSM, partly in the interests of economy, but 
also because it is hard to conceive of the hidden sector fields as 
carrying the global, but not the local, quantum numbers of the SM 
sector.

Once we have generated the soft SUSY-breaking terms at a high scale, we 
have a theory where sparticle masses of a given spin are universal. It 
is usual to identify this scale with the scale for grand unification 
(GUT scale). The mass spectrum of the model at the electroweak scale is 
now generated by allowing the masses and couplings to run down from the 
GUT to the electroweak scale using the renormalisation group equations 
(RGE) for the model. Some judicious manipulation of independent 
parameters replaces the free parameter $\mu$ by its sign, and the ratio 
of vacuum expectation values (VEVs) of the two Higgs boson doublets (at 
the electroweak scale), which is denoted by $\tan\beta$. This is a 
natural parametrisation that ensures that one of the Higgs mass 
parameters is driven negative at precisely the electroweak scale -- 
which provides an `explanation' of the phenomenon of electroweak 
symmetry-breaking (EWSB).

Thus the cMSSM has several merits. It does not require much input from 
the hidden sector except supersymmetry-breaking, but we can generate the 
entire spectrum of masses, couplings and mixing angles in terms of just 
four parameters and a sign, viz. $m_0$, $m_{1/2}$, $A_0$, $\tan\beta$ 
and the sign of $\mu$. All that one has to do then, is to choose a point 
in this limited parameter space, and the entire gamut of SUSY 
phenomenology is uniquely determined by this choice. For this reason, 
and for the theoretical niceties described above, the cMSSM has been the 
favourite choice for studies of new physics beyond the SM.

In recent times, we have seen a breaking-away from the cMSSM paradigm, 
with the introduction of several non-universal versions of SUSY 
\cite{nususy}. Though these offer a richer phenomenological fare --- 
which is to be expected when the number of free parameters increases --- 
all these models have, implicitly or otherwise, to make some assumptions 
about the hidden sector which the cMSSM does not. This is not to say 
that non-universal models should not be investigated, but it should be 
clear that from the point of view of economics and aesthetics, that the 
cMSSM has definite advantages.

Given the above, it is not surprising that the cMSSM is often given top 
priority whenever experimental searches for new physics beyond the SM 
are considered. This is indeed so, and has been the case for the past 
few decades. Of course, all such searches till date have yielded 
negative results and only succeeded in ruling out parts of the cMSSM 
parameter space. It is important, therefore, to see how these 
restrictions arise, and how seriously one should take them. Constraints 
on the cMSSM arise from four main sources. These are listed below.

\vspace*{-0.2in}

\begin{itemize}

\item {\sl Theoretical Considerations}: Except for rather loose 
naturalness considerations, there are no {\it a priori} theoretical 
guidelines for the choice at the GUT scale of values of $m_0$, $m_{1/2}$ 
and $A_0$, or the sign of $\mu$. For $\tan\beta$, however, we note that 
the Yukawa couplings of the top and bottom quarks remain comfortably 
perturbative so long as $1.2 \lsim \tan\beta \lsim 65$ \cite{martin}.

Very small values of $m_0$ and $m_{1/2}$ ($\sim$ few GeV) are not viable 
in the cMSSM, for then the RGE would drive the electroweak 
symmetry-breaking to happen rather close to the GUT scale, and this 
would imply a much lower GUT scale than appears to be indicated by the 
measured running of the gauge coupling constants. For larger values of 
$m_0$ and $m_{1/2}$, there arise two kinds of {\it a posteriori} constraints 
which act collectively on the parameters $m_0$, $m_{1/2}$ and $A_0$ when 
the cMSSM spectrum is run down from the GUT scale to the electroweak 
scale. One is the requirement that the scalar potential in the theory 
remain bounded from below -- this is referred to as the {\it vacuum 
stability} constraint \cite{djouadi}. The other is the requirement that 
the lightest SUSY particle (LSP) be a neutral particle -- which is 
demanded if it is to be the major component of dark matter. This is 
found to rule out a region of the parameter space where the RGE 
evolution makes the lighter stau $\widetilde{\tau}_1$ the LSP.

Another consideration, which is not a constraint but may be regarded as 
some sort of wishful thinking, is a requirement that the parameters 
$m_0$, $m_{1/2}$ and $A_0$ not be much more than a few TeV. This is 
because higher values of these parameters -- especially the first two -- 
tend to drive the masses of all the SUSY particles outside the kinematic 
range of the LHC (and even its foreseeable successors), while the 
lightest Higgs boson mass gets pushed close to a value around 120~GeV. 
In this, so-called {\it decoupling limit} the cMSSM Higgs boson would be 
indistinguishable, for all practical purposes, from its SM counterpart. 
Such a scenario, though by no means impossible, would be a great 
disappointment for seekers of new physics, as it would leave the 
existence of SUSY as a wide open question without a hope of solution in 
the near future. Of course, requiring the $m_0$, $m_{1/2}$ and $A_0$ to 
be in this convenient range is essentially dogma, but it is what renders 
studies of the present kind worth carrying out.

\item {\sl Indirect effects at low-energy experiments}: The cMSSM can 
affect low-energy processes when the relatively-heavy superparticles 
appear in Feynman diagrams {\rm at the loop level}. Any measurement of a 
low-energy process which ($a$)~get contributions from these particles, 
and ($b$)~is measured with sufficient accuracy to access these usually 
small effects, will impose a constraint on the cMSSM. Based on these two 
criteria, we can now discern {\it three} distinct types of low-energy 
constraints.

\begin{enumerate}

\item The first type is where the low-energy effect is observed and 
measurement is consistent with the SM prediction. In this case, any 
contributions from the cMSSM will have to be small enough to fit into 
the small leeway allowed by the error bars. Such constraints have a 
tendency to get tighter and tighter as more data are collected in an 
experiment and the error bars shrink. The relevant example of this is 
the radiative $B$ decay $B \to X_s \gamma$, where the measured value of 
the branching ratio ${\rm BR}(B \to X_s \gamma)$ is $(3.55 \pm 0.24 \pm 
0.09) \times 10^{-4}$\cite{Asner:2010qj} against an SM prediction of 
$(3.15 \pm 0.23) \times 10^{-4}$ \cite{Misiak:2006zs}. Thus, including 
the intrinsic cMSSM uncertainty of about $0.15 \times 10^{-4}$ as given 
in \cite{Ellis:2007fu} (which was based on \cite{Degrassi:2000qf}, we 
can set the 95\% allowed range to be $(3.55 \pm 0.95) \times 10^{-4}$. 
This means that the cMSSM contribution must satisfy
$$
-0.55 \times 10^{-4} \leq {\rm BR}_{\rm cMSSM}(B \to X_s \gamma) \leq 1.35 
\times 10^{-4} \ .
$$

\item The second type is where the SM effect is smaller than the 
existing experimental upper bound, which leaves room for reasonably 
large contributions from the cMSSM. Even more than the previous case, 
these constraints get tighter as the lower bound is tightened, but 
often, even with improvements in experimental techniques, this bound 
remains significantly above the SM prediction, so that there is always 
some room for a cMSSM contribution. A good example of this is the decay 
$B_s \to \mu^+ \mu^-$, where the experimental upper bound on ${\rm 
BR}(B_s \to \mu^+ \mu^-)$ has been recently improved to $4.5 \times 
10^{-9}$ \cite{Aaij:2012ac} against a SM prediction of $(3.2 \pm 0.2) 
\times 10^{-9}$ \cite{Buras:2010wr}. In order to take into account the 
theoretical uncertainties, in our numerical analysis we set the 95\% 
upper limit to be $5.0 \times 10^{-9}$. Thus, the cMSSM contribution 
must satisfy
$$
{\rm BR}_{\rm cMSSM}(B_s \to \mu^+ \mu^-) \leq 1.8 \times 10^{-9} \ .
$$

\item There exists a third -- and rare -- type of low-energy process 
where the experimental result is not consistent with the SM prediction 
at some level between 1--2$\sigma$. The experience of the past few 
decades has generally been that a more accurate measurement of the 
process, or a more sophisticated computation of the SM prediction 
generally brings the two into perfect consistency, but there are two 
results which have till date defied this comfortable precedent. One is 
the measurement of the anomalous magnetic moment $(g-2)_\mu/2$ of the 
muon \cite{Bennett:2006fi,Davier:2010nc}, and the other is the decay 
width for $B^+ \to \tau^+ \nu_\tau$ \cite{Aubert:2008ac}. In the former 
case, the SM prediction is too small to explain the experimental data at 
the level of more than 3$\sigma$. For $\mu < 0$, the cMSSM contributions 
tend to {\it decrease} the SM contribution still further, so that the 
constraints from this process become very strong in this regime. In the 
case of $B^+ \to \tau^+ \nu_\tau$, too, the SM prediction is again too 
low at the level of around 2$\sigma$, and once again the cMSSM 
contributions tend to push the predicted value down rather than up. In 
fact, if these two anomalies are taken together and at face value, very 
little of the cMSSM parameter space is still viable 
\cite{Bhattacherjee:2010ju}, and this little bit can barely survive the 
experimental constraints already available. However, one can argue that 
if we are to take these discrepancies at face value, then the SM is also 
ruled out and hence we have already found new physics! In view of the 
uncertainties in the SM prediction for these processes 
\cite{Lunghi:2011xy}, particularly for $(g-2)_\mu/2$, this is surely too 
bold a prediction to make at this stage. Hence, if one shrinks from 
declaring the discovery of new physics in this context, then, by the 
same token, one must also refrain from declaring the imminent demise of 
the cMSSM in the same context. \end{enumerate}

A generic feature of low-energy constraints is that the restrictions 
obtained from them always assume that the cMSSM (for example) is the 
only new physics contribution to the relevant process. Obviously, if 
there are other new physics contributions, the constraints on the cMSSM 
would change, being either strengthened or relaxed, depending on the 
relative sign of the two new physics contributions. Of course, one can 
postulate such effects ad infinitum, and therefore, the usual practice 
is to invoke Occam's razor and demand that we consider only one sort of 
new physics at a time. But Nature may well be perverse in this respect, 
and hence, there is always an element of wishful thinking when we apply 
any indirect constraints. Ultimately, indirect constraints only become 
really strong when backed up by direct evidence.

\item {\sl Direct searches at high energy colliders}: Like any model of 
new physics, the cMSSM makes predictions of new particles and 
interactions, and when one performs a direct search for these in a high 
energy experiment without observing any of the predicted effects, one 
ends up ruling out the part of the parameter space.

One of the important features of the cMSSM is that the LSP is always the 
lightest neutralino $\widetilde{\chi}_1^0$ and all SUSY particles 
undergo cascade decays with the $\widetilde{\chi}_1^0$ as the final 
product. Guaranteed stability by the conservation of $R$-parity -- a 
crucial feature of the cMSSM -- this LSP interacts very weakly with the 
matter in the detectors and usually flies away to add a minuscule amount 
to the dark matter component of the Universe, leaving a momentum 
imbalance in the observed events, which is measured by the missing 
transverse momentum (MpT) or its equivalent, the missing transverse 
energy (MET). This missing energy provides a unique way to search for 
signals of the cMSSM, for any process at a high energy collider which 
produces a pair of SUSY particles (the pair being required by $R$-parity 
conservation) will be followed by direct or cascading decays of these 
SUSY particles, always ending up in a set of SM particles and large MET.

At the LHC, in particular, the most viable signals will arise when the 
strongly- interacting SUSY particles, viz. the squarks and the gluinos, 
are produced in pairs and then each member of the pair decays to other 
strongly-interacting particles and the LSP, leading to a final state 
with multiple jets and substantial MET. Searches for such states form 
the spearhead of the cMSSM effort at the LHC, and also yield the 
strongest constraints from direct searches. We shall have more to say on 
this subject presently.

It is important to note that $R$-parity conservation is not demanded by 
any of the underlying symmetries of the cMSSM, and has to be imposed by 
hand. This may be regarded as another element of wishful thinking in the 
model. Of course, models with non-conservation of R-parity must violate 
either lepton number or baryon number and hence, will have their own set 
of distinctive signals.

\item {\sl Dark matter requirements}: Though SUSY was originally invoked 
to solve the gauge hierarchy-cum-fine tuning problem in the SM, that 
`motivation' can be argued away if one does not believe in a GUT scale 
below the Planck scale. However, a more attractive feature of the cMSSM 
is that the LSP -- which is stable and has weak interactions with matter 
-- is just the sort of weakly-interacting massive particle (WIMP) which 
cosmologists require in order to understand the known behaviour of 
cosmic dark matter. Here we may recall the two essential features of 
dark matter. In the first place, it is `dark', i.e. it does not radiate, 
exactly as a relic density of massive neutralinos might be expected to 
do. Secondly, it is composed of non-baryonic gravitating matter -- 
probably WIMPs -- as is proved rather spectacularly by studies of the 
Bullet Cluster and similar objects in the Universe.

Purely by studying rotation curves of galaxies and by collecting and 
collating information about gravitational lensing, one can establish 
that the amount of dark matter in the Universe is definitely several 
times more than the amount of visible, baryonic matter. However, the 
tiny acceleration of the Universe requires a delicate balance between 
the attractive force of gravitating dark matter and the mysterious 
repulsive force of dark energy. In fact, to have the working model of 
inflationary cosmology which will explain the highly-precise Planck data 
on the cosmic microwave background radiation (CMBR), one requires a 
rather fine-tuned relic density $\Omega_d h^2$ of dark matter. The 
specific requirement is that \cite{Komatsu:2010fb}
$$
\Omega_d h^2 \simeq 0.1123 \pm 0.0035 
$$
at 95\% C.L. In turn, this imposes constraints on the cMSSM, since all 
points in the parameter space are not consistent with such a fine- tuned 
relic density. In the cMSSM, one would require a fine balance between 
processes which produce neutralino pairs, and co-annihilation processes 
where a neutralino pair interacts to form ordinary matter.

Apart from the inferences from the dark matter relic density, one can also
impose constraints on the parameter space of any model like the cMSSM
using the data from direct and indirect searches for dark matter at 
terrestrial and satellite-based experiments. A comprehensive summary
of the results of direct searches can be found in Figure~13 of Ref.~\cite{CRESST}. Unfortunately, most of the results from different
experiments are, at the moment, not even compatible with each other. 
It has also been speculated that there may be large uncertainties in the 
cross-sections of dark matter particles with nucleons \cite{lattice}.
In view of these, we feel that it is premature to use these results 
to constrain the cMSSM parameter space. Constraints from indirect 
experiments are rather weak \cite{Kersten}, and do not affect our 
results in any significant way.   

It remains, then, to ask how seriously one should take any of the dark matter 
constraints on the cMSSM. The status of this is really somewhat like the 
indirect constraints coming from low energy. For after all, dark matter 
could be some other form of matter, or even a mixture of neutralinos 
with some other invisible particles. As before, one can invoke Occam's 
razor to say that one should consider only single-component dark matter, 
but in the final reckoning that is mere dogma, and the ultimate test 
will have to come from a direct discovery. Furthermore, the LSP is 
stable only if $R$-parity is conserved, and this, as we have seen, is an 
ad hoc assumption.

\end{itemize}

\vspace*{-0.2in}
Thus, the cMSSM is constrained in four different ways, each process 
ruling out some small part of the parameter space, subject to caveats as 
mentioned above. Taking all factors into consideration, we now make a 
numerical analysis of the allowed cMSSM parameter space with the 
following constraints.

\vspace*{-0.2in}

\begin{enumerate}
\item The ranges of the cMSSM parameters are:
$$
0 \leq m_0 \leq 4~{\rm TeV} \qquad
0 \leq m_{1/2} \leq 1~{\rm TeV} \qquad
-1~{\rm TeV} \leq A_0 \leq +1~{\rm TeV} \qquad
1 \leq \tan\beta \leq 60
$$
with $\mu > 0$, since $\mu < 0$ is strongly disfavoured not only by 
$(g-2)_\mu/2$ by also by $B \to X_s \gamma$.
\item For all choices of the parameters the scalar potential must be 
bounded from below and the LSP must not be the $\widetilde{\tau}_1$ (or 
any other charged particle). These constraints have, of course, been 
imposed ever since the cMSSM was proposed.
\item The entire set of constraints from the CERN LEP-2 collider data is 
imposed. The most restrictive among these are the requirements that 
\begin{itemize}
\item the mass of the lighter chargino must satisfy 
$M(\widetilde{\chi}_1^\pm) \geq 94.0$~GeV \cite{pdg}, and
\item the mass of the lightest Higgs boson $m_h \geq 93.0$~GeV for 
$\tan\beta \gsim 6$, and $m_h \geq 114.0$~GeV \cite{lep} for $\tan\beta 
\lsim 6$, with a range of intermediate values in the neighbourhood of 
$\tan\beta \simeq 6$.
\end{itemize}
Constraints arising from other considerations (such as, for example, the 
mass of the lighter stop and the lighter stau) are generally subsumed in 
the disallowed parameter space due to these two major constraints, but 
we impose them nevertheless. Like the previous case, these constraints 
have been in place for some time now, ever since the final data analyses 
from the LEP Electroweak Working Group became available.

\item The area of parameter space disallowed by direct searches at the 
LHC is taken over from the ATLAS and CMS Collaborations 
\cite{atlaslimit,cmslimit}. These are obtained by combining the negative 
results of searches in many channels, but the most important of these is 
the search in the jets + MET channel.

\item Constraints on the cMSSM from the rare decay $B \to X_s \gamma$ 
\cite{Asner:2010qj} are imposed. The specific requirement is that the 
cMSSM contribution to the decay width should satisfy
$$
-0.55 \times 10^{-4} \leq {\rm BR}_{\rm cMSSM}(B \to X_s \gamma) \leq 1.35 
\times 10^{-4} \ ,
$$
at 95\% C.L.. 

\item Constraints on the cMSSM from the recently-measured upper bound on 
the rare decay $B_s \to \mu^+\mu^-$ \cite{Aaij:2012ac} are imposed. This 
measurement has recently been substantially improved by the LHCb 
experiment and their updated result has been used in this work.The 
specific requirement is that the cMSSM contribution to the decay width 
should satisfy
$$
{\rm BR}_{\rm cMSSM}(B_s \to \mu^+ \mu^-) \leq 1.8 \times 10^{-9} \ ,
$$ 
at 95\% C.L..

\item Finally, we have mapped the part of the cMSSM parameter space 
which is compatible with the requirement that the observed dark matter 
component of the Universe be purely a relic density of LSPs 
$\widetilde{\chi}_1^0$. The specific requirement is that
$$
0.1053   \leq \Omega_d h^2 \leq 0.1193
$$
at 95\% C.L. \cite{Komatsu:2010fb} and the stable value of $\Omega_d 
h^2$ is calculated by solving the relevant Boltzmann equation for the 
time evolution of the relic density. We do not treat this as a 
constraint, but merely show the allowed region alongside that permitted 
by all other considerations.

\end{enumerate}

\vspace*{-0.2in} 

Apart from the requirement that $\mu > 0$, we have not imposed any 
specific constraint from the data on $(g-2)_\mu/2$, and we have chosen 
not to consider low-energy constraints arising from $B^+ \to \tau^+ 
\nu_\tau$. This last has not been taken into account because we feel 
that the situation vis-a-vis the SM has not yet stabilised and it may be 
premature to use this to constrain new physics. But all this is not to 
say that other constraints on the cMSSM parameter space from low-energy 
data do not exist --- in fact, every measurement which is compatible 
with the SM prediction and has a cMSSM contribution will impose a 
constraint. However, we find that, except for $(g-2)_\mu/2$ and $B^+ \to 
\tau^+ \nu_\tau$, none of these are as restrictive on the cMSSM 
parameter space as the set of constraints listed above: the range of 
parameter space affected by these is always a subset of that ruled out 
by the combination of those from the above-listed set.
  
At this juncture, we note that several papers\cite{bsmumuimpli} have 
appeared in which the constraints on the cMSSM from the decays 
$B \to X_s \gamma$ and $B_s \to \mu^+\mu^-$ have been studied, both 
independently, and in conjunction with the LHC constraints from direct 
searches. Some of these have been used to predict the most likely values 
of the cMSSM parameters. We have chosen the more conservative approach
of mapping out the parts of the parameter space which are disallowed,
and assuming equal {\it a priori} probability for the rest. Our presentation 
of the constraints is, therefore, very close to the way in
which direct constraints from the experimental data are available.

We are now in a position to consider the cMSSM parameter space and see 
how it gets constrained when the above conditions are applied. Our 
detailed results are given in the next section.

\vspace*{-0.2in} 

\section{Update on the cMSSM Parameter Space}

\vskip -10pt
Once we have fixed the sign of $\mu$ to be positive, as explained above, 
the cMSSM parameter space is a four-dimensional space, with the 
parameters being $m_0$, $m_{1/2}$, $A_0$ and $\tan\beta$, as described 
above. Since one can plot only two of them at a time, it is traditional 
to pick two of these parameters and keep the others either fixed, or 
floating. The most common plots are made in the $m_{1/2}$--$m_0$ plane, 
with $A_0$ and $\tan\beta$ fixed. This is because the masses of the 
superparticles depend most directly on these two parameters $m_0$ and 
$m_{1/2}$, with the other two contributing mostly through mixing that 
occurs between pure superparticle states when the electroweak symmetry 
is broken. For our first plot, therefore, we choose the $m_{1/2}$--$m_0$ 
plane, for three separate values of $A_0 = 0$ and $\pm 1$~TeV, with 
$\tan\beta = 10$, the last choice being influenced by the latest plots 
available from the ATLAS and CMS Collaborations. We shall later have 
occasion to vary $A_0$ and $\tan\beta$, so this particular choice may be 
regarded merely as an opening gambit. We generate the cMSSM spectrum 
using the software {\sc SuSpect} \cite{suspect} and calculate the 
low-energy observables (including the dark matter relic density) using 
{\sc SuperISO} \cite{Arbey:2009gu}. Our results are shown in 
Figure~\ref{fig:mZmH_10}. It may be noted that these plots correspond to 
a top quark mass of 172.9~GeV\cite{pdg}.

The three panels in Figure~\ref{fig:mZmH_10} correspond, from left to 
right, to choices of $A_0 = +1$~TeV, 0 and $-1$~TeV respectively. In 
each panel, we have plotted $m_{1/2}$ in the range $0-1$~TeV, and $m_0$ 
in the range $0-4$~TeV, keeping $\tan\beta = 10$ as mentioned above. It 
is worth recalling, at this juncture, that the superparticle masses tend 
to grow with both $m_0$ and $m_{1/2}$, and hence, may roughly be said to 
grow along the north-east diagonal of the plot. Thus increased energy of 
the machine will increase the discovery reach approximately diagonally 
along these plots, and this is what is, in fact seen.

\begin{figure}[htb]
\setcounter{figure}{0}
\centerline{ \epsfxsize= 6.3 in \epsfysize= 2.8in \epsfbox{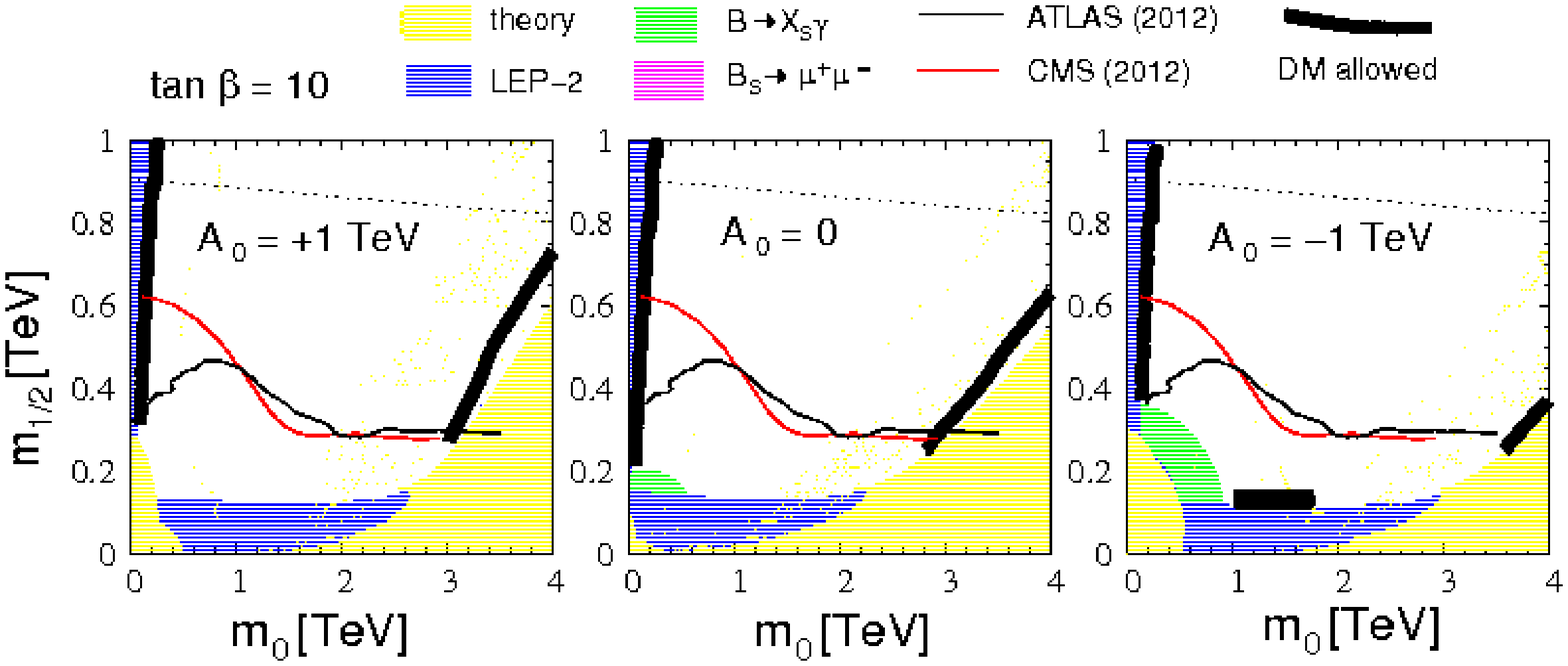} }
\vskip -10pt
\def\baselinestretch{0.8}

\caption{{\footnotesize Illustrating constraints on the $m_0$--$m_{1/2}$ 
plane in the cMSSM for $\tan\beta = 10$, as well as the region which 
explains the observed relic density of dark matter. The details are 
marked on the different panels or in the key above. The dotted line is 
the contour for a gluino mass of 2~TeV. Note that $\mu > 0$ for all the 
plots. The ATLAS and CMS exclusion curves correspond to $A_0 = 0$ but 
are not very sensitive to $A_0$ or even $\tan\beta$.
}}
\def\baselinestretch{1.2}
\label{fig:mZmH_10}
\end{figure}

In each plot, the region shaded yellow is ruled out by `theory 
constraints'. Of these, the requirement of vacuum stability is the 
dominant constraint in the region close to the abscissa and the stau-LSP 
is the dominant contribution in the region close to the ordinate. For 
large values of $m_0$ some of the points are disallowed simply because 
the renormalisation group equations (RGE) used to calculate the cMSSM 
spectrum at the electroweak scale have no real solutions. Shapes vary 
somewhat between the three panels, illustrating the influence of the 
parameter $A_0$ on the RGE running of the cMSSM parameters, but the 
basic features are common, with small values of $m_0$ and $m_{1/2}$ 
being ruled out in every case. The regions shaded blue in the three 
panels of Figure~\ref{fig:mZmH_10} correspond to constraints arising 
from LEP-2 data. These are generally stronger than the theoretical 
constraints, except for extreme values of $m_{1/2}$. Most of the LEP-2 
disallowed region arises from the chargino mass constraint. The small 
sliver of space ruled out by LEP-2 for very low values of $m_0$ at 
relatively large values of $m_{1/2}$ corresponds to negative searches 
for light stau states at LEP-2.

In Figure~\ref{fig:mZmH_10}, constraints arising from the non-discovery 
of cMSSM signals in 4.7~fb$^{-1}$ of data at the ATLAS (CMS) detector 
are shown by the solid black (red) line, with the region {\it below} the 
curve getting ruled out. The ATLAS exclusion curve arises from a 
combination of all processes, whereas the CMS exclusion plot arises only 
from searches for the 0 lepton + jets + MET final states made using 
`razor variables' \cite{cmslimit}. These published analyses both choose 
$A_0 = 0$. Strictly speaking, therefore, this constraint should appear 
only in the central panel. However, the constraints from a jets + MET 
search are not very sensitive to the choice of $A_0$, and hence, we have 
made bold to use the same curve for all the three cases. Differences, if 
any, will be marginal, and should not make any qualitative impact on our 
discussions regarding these plots. The most important qualitative 
feature of these constraints is that, unless $A_0$ is strongly negative, 
they represent significant improvements over the LEP-2 bounds. As more 
data is collected and analysed, one may expect the LHC constraints to 
become stronger, and eventually cover most of the parameter space marked 
in the panels of Figure~\ref{fig:mZmH_10}. It may be noted that though 
we have not marked any projected reach of the LHC on these plots, a 
ballpark estimate may be formed from the contour of gluino mass 2~TeV, 
which is shown by the dotted line near the top of each panel. Therefore, 
we may conclude that eventually the LHC will be able to explore 80--90\% 
of the parameter space shown in Figure~\ref{fig:mZmH_10}, barring the 
uppermost regions of each panel. Of this, roughly one half is already 
ruled out, but this is equivalent to saying that roughly one half is 
still allowed.

The constraints from low-energy data are marked on the graph in green 
for $B \to X_s \gamma$ and pink for $B_s \to \mu^+\mu^-$. What 
immediately strikes the eye is the fact that these are rather weak -- at 
least, in the three panels of Figure~\ref{fig:mZmH_10} --- where the 
strongly $\tan\beta$ dependent $B_s \to \mu^+\mu^-$ constraint (see 
below) makes no appearance at all, while the $B \to X_s \gamma$ data 
adds on a little to the LEP-2 constraint for $A_0 = -1$~TeV. Even this 
is totally subsumed in the LHC constraints. One may be tempted to 
conclude that low-energy measurements are not competitive with the 
direct searches in constraining the cMSSM parameter space, but we must 
remember that the plots of Figure~\ref{fig:mZmH_10} are for a fixed 
value of $\tan\beta$. The situation changes, quite dramatically, when we 
go to larger values of $\tan\beta$.

Before we go on to discuss high $\tan\beta$ results, however, let us 
note that the regions in Figure~\ref{fig:mZmH_10} which are consistent 
with the dark matter (DM) relic density are marked by narrow black bands 
on all the three plots. This allowed band appears only as the so-called 
`stau co-annihilation region', i.e. very close to the region disallowed 
by the stau-LSP constraint, and again for large values of $m_0$, in the 
so-called `funnel region'. Nevertheless, it is heartening to see that 
there is always a region of the parameter space which can be the 
explanation of the dark matter phenomenon in the cMSSM. This model has 
not, therefore, lost its most attractive phenomenological feature, and 
the continuation of at least one small portion of the black bands into 
the regions inaccessible to the LHC tells us that even if the LHC 
completes its run without finding any signatures of the cMSSM, we will 
still be able to argue that the neutralino (albeit a heavier one than we 
now think) is the main component of the dark matter.

To sum up this part of the discussion, then, {\it for low values of} 
$\tan\beta$, for which the value $\tan\beta = 10$ serves as a benchmark, 
the cMSSM is under no serious threat (unless, as we have seen, 
constraints from $(g-2)_\mu/2$ and $B^+ \to \tau^+\nu_\tau$ are imposed
 \cite{Bhattacherjee:2010ju}) from a combination of low-energy data and 
direct searches. Even if the next round of direct searches throws up a 
negative result, constraining the parameter space still further, it 
should not be regarded as the death-knell of the cMSSM, for all that may 
be happening is that we are in the process of eliminating a barren 
region in the parameter space as we approach the actual region of 
interest.

\begin{figure}[htb]
\centerline{ \epsfxsize= 6.5 in \epsfysize= 4.8in \epsfbox{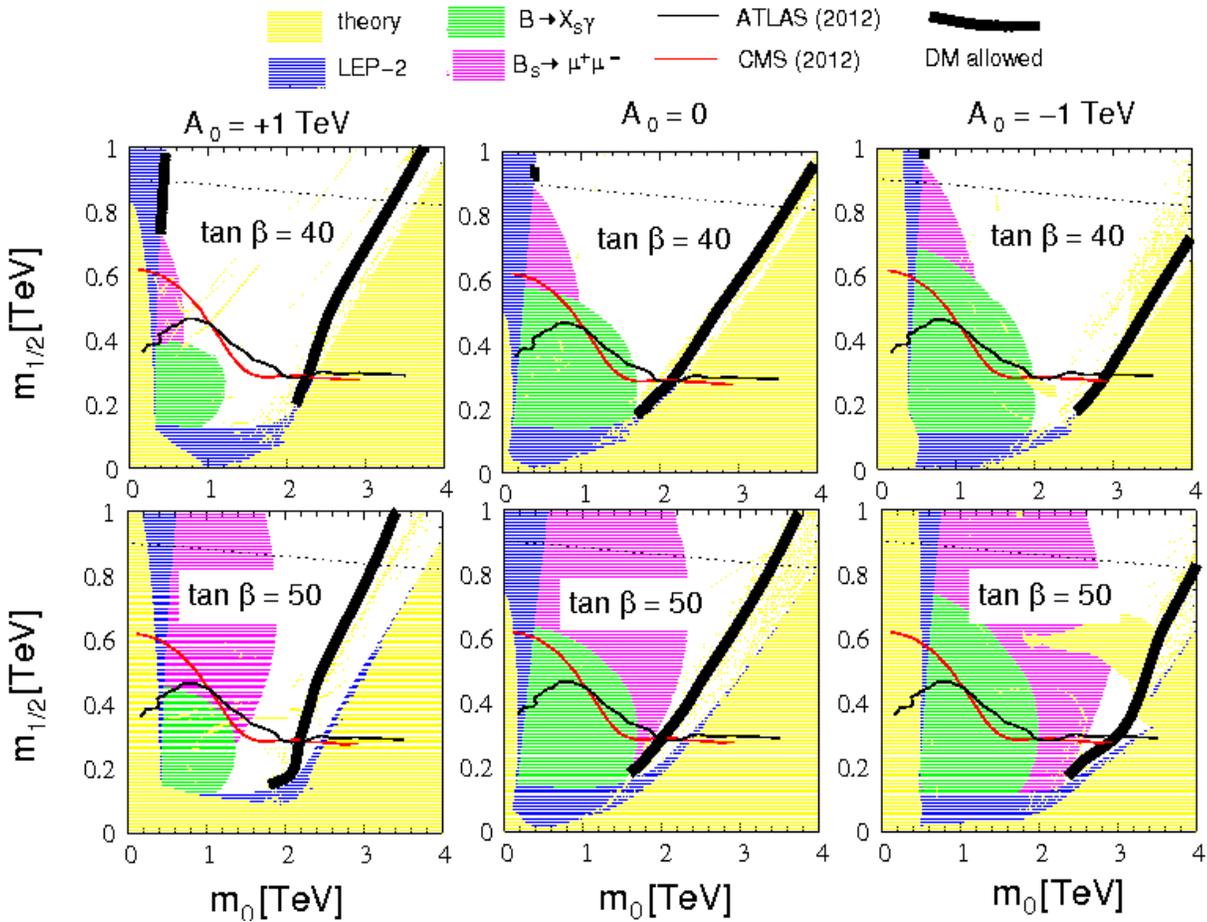} }
\vskip -10pt
\def\baselinestretch{0.8}
\caption{{\footnotesize Illustrating constraints on the $m_0$--$m_{1/2}$ 
plane in the cMSSM for high values $\tan\beta = 40, 50$. All notations 
and conventions are the same as in Figure~\ref{fig:mZmH_10}. In these 
plots $B$-physics constraints become significant, especially for $A_0 
\leq 0$. Note that the dark matter-compatible band always lies in the 
allowed region. Some of the (yellow) islands indicated as theory-disallowed
for large values of $m_0$ represent numerical instabilities in the spectrum
generator {\sc SusPect}. }}
\def\baselinestretch{1.2}
\label{fig:mZmH}
\end{figure}

Let us now see what happens when $\tan\beta$ is increased. We have seen 
that for $\tan\beta = 10$, the only effect of including low-energy 
constraints is to marginally extend the LEP-2 bound \cite{lep}, and that 
too, only for $A_0 = -1$~TeV. This feature continues to hold all the way 
up to for $\tan\beta = 35$, which covers a substantial fraction of the 
theoretically allowed range, (viz. up to 60). Around $\tan\beta = 40$, 
however, the low-energy constraints begin to become significant, and for 
$\tan\beta = 50$, they outstrip the direct searches and constrain a 
significant extra part of the parameter space. This is illustrated in 
Figure~\ref{fig:mZmH}.

The conventions followed in Figure~\ref{fig:mZmH} are exactly the same 
as those followed in Figure~\ref{fig:mZmH_10}, and are indicated, as in 
Figure~\ref{fig:mZmH}, by the little key on the top. The three panels in 
the first row correspond to $\tan\beta = 40$, while the three panels in 
the second row correspond to $\tan\beta = 50$. In each row, the three 
panels correspond to choices of $A_0 = +1$~TeV, 0 and $-1$~TeV (from 
left to right). Regions shaded yellow (blue) correspond to constraints 
from theory (LEP-2), and solid lines marked in black (red) correspond to 
the exclusion plot of the ATLAS (CMS) Collaboration\footnote{We 
reproduce the exclusion plots already exhibited in 
Figure~\ref{fig:mZmH_10}, which, strictly speaking, are valid only for 
$\tan\beta = 10$~GeV. However, as we have taken the combined exclusion 
plot from ATLAS and the purely hadronic exclusion plot from CMS, the 
larger values of $\tan\beta$ in these plots will not make a significant 
difference.}. The black strips correspond to regions which are 
consistent with the neutralino interpretation of dark matter, and the 
dotted line near the top of each panel corresponds to the contour of 
$M(\tilde{g}) = 2$~TeV. For large values of $m_0$, some of the yellow 
islands indicating theory-disallowed regions, especially in the bottom
right panel, represent numerical instabilities in the spectrum
generator {\sc SusPect}, and would be allowed if a different spectrum
generator had been used.  

Let us begin by discussing the situation for $\tan\beta = 40$, i.e. the 
upper row of panels in Figure~\ref{fig:mZmH}. The first thing that 
strikes the eye is that the theoretically constrained area is larger 
than in the case of $\tan\beta = 10$, not only in the region which is 
identified as due to a stau LSP, but over a very large region for higher 
values of $m_0$. The first part is easy to understand, since the 
off-diagonal terms in the mass matrix for staus $\widetilde{\tau}_1$, 
$\widetilde{\tau}_2$ are proportional to $\tan\beta$. Larger values of 
$\tan\beta$ can be interpreted as causing a larger splitting between the 
mass of the heavier and lighter stau, thus pushing the mass of the 
lighter stau $\widetilde{\tau}_1$ downwards, below the mass of the 
neutralino $\widetilde{\chi}^0_1$. This last is not much changed by 
increasing $\tan\beta$ --- a statement which is generically true for all 
gauginos, including the lighter chargino $\widetilde{\chi}^\pm_1$, as a 
result of which the constrained region from the LEP-2 data remains much 
the same as before. For large values of $m_0$, the RGE evolution is 
simply not enough to drive one of the scalar mass parameters to negative 
values, and this manifests as non-convergence of the RGE when we demand 
such negative values. Alternatively we can simply say that for such 
parameter choices the electroweak symmetry remains unbroken. Even more 
than the theoretical constraints, however, for large values of 
$\tan\beta$ the constraints from low-energy measurements become much 
more significant. For example, the constraints from $B \to X_s \gamma$, 
which made such a modest appearance in the case of $\tan\beta = 10$, now 
begin to outstrip the LHC exclusion boundaries, especially for $A_0 \leq 
0$. Even more dramatic than the growth of the $B \to X_s \gamma$ 
constrained region is the appearance of a significant (pink-shaded) 
region which is now disallowed by the $B_s \to \mu^+\mu^-$ constraint. 
For $\tan\beta = 40$, this is still a smallish appendage to the region 
already disallowed by other constraints, but if we now look at the lower 
set of three panels in Figure~\ref{fig:mZmH}, where $\tan \beta = 50$, 
it is clear that this new constraint affects large parts of the 
parameter space which are allowed by all other constraints. This growth 
in importance of the $B_s \to \mu^+\mu^-$ constraint can be readily 
understood in terms of an enhanced cMSSM contribution from the lighter 
stop $\widetilde{t}_1$. As in the case of staus, larger values of 
$\tan\beta$ increase the diagonal terms in the stop mass matrix and 
drive the mass eigenvalue corresponding to the $\widetilde{t}_1$ to 
lower values. Indeed, for large $\tan\beta$, the cMSSM contribution is 
known \cite{deboer} to scale as $\tan^6 \beta/(M_{H^+}^2 - M_W^2)^2$.

If we take a quick glance at the lower three panels in 
Figure~\ref{fig:mZmH}, one might be tempted to say that a value of 
$\tan\beta$ as large as 50 seems to be disfavoured because the 
low-energy constraints combine with the existing ones from theory and 
direct searches to choke off most of the parameter space accessible to 
the LHC. However, large $\tan\beta$ values are interesting as they led 
to distinctive sparticle decay signatures, especially those involving 
tau final states. It is apparent from the very same figure that the 
black bands, denoting consistency with the dark matter relic density, go 
right through the allowed `funnel region' in every panel, showing that a 
high $\tan\beta$ solution of the dark matter problem is very much a 
viable one.

Given that the $m_0$--$m_{1/2}$ plane is so much more constrained, all 
things taken together, in the high $\tan\beta$ cases, it is natural to 
ask how accessible the allowed parameter space will be to the LHC runs 
in the near and more distant future.  Of course, once these runs are 
over, we may expect proper exclusion plots from both ATLAS and CMS 
Collaborations, but a certain amount of curiosity about what could be 
the outcome of these runs is inevitable. To satisfy this, we have chosen 
four benchmark points, within the region mapped in 
Figure~\ref{fig:mZmH}, each of which is just allowed by the present 
constraints and two of which are compatible with the dark matter relic 
density. The exact choices of parameters corresponding to these points 
are given in Table~\ref{tab:ABCD} below. In the next section we shall 
take up a more detailed study of cMSSM signals at the LHC for these 
points.

\begin{table}[htb]
\begin{center}
\begin{tabular}{crrrrr}
\hline
BP & $m_0$ & $m_{1/2}$ & $A_{0}$ & $\tan\beta$  & sgn~$\mu$ \\  
\hline\hline
A  & 0.4  & 0.75      &  ~1.0       &    40.0        &  +1       \\  
B  & 2.4  & 0.38      &  ~0.0       &    40.0        &  +1       \\
C  & 1.0  & 0.55      &  ~0.0       &    40.0        &  +1       \\
D  & 2.0  & 0.35      &  -1.0        &    40.0        &  +1       \\
\hline
\end{tabular}
\caption{\footnotesize The cMSSM parameters for the four benchmark 
points (BP) chosen for a detailed collider analysis in the context of 
the current LHC run. The first three columns are in units of TeV. All 
the points are consistent with the experimental constraints. The points 
A and B are compatible with neutralino dark matter; C and D are not.}
\label{tab:ABCD}
\end{center}
\end{table}

\vspace*{-0.2in}
Before taking up detailed collider studies, however, it is instructive 
to study the cMSSM parameter space in some other ways. For example, 
since $m_0$ and $m_{1/2}$ are not the parameters directly measurable at 
the LHC, one might ask, instead: what are the constraints on the masses 
of the squarks and gluinos? Here it should be noted that while all the 
eight gluinos have the same mass $M_{\widetilde{g}}$, this is not true 
of the twelve squarks. This is especially true of the third generation; 
the squarks of the first two generations are practically degenerate, 
with $M_{\widetilde{q}}$ denoting their common mass.

\begin{figure}[ht]
\centerline{ \epsfxsize= 6.5 in \epsfysize= 3.2in \epsfbox{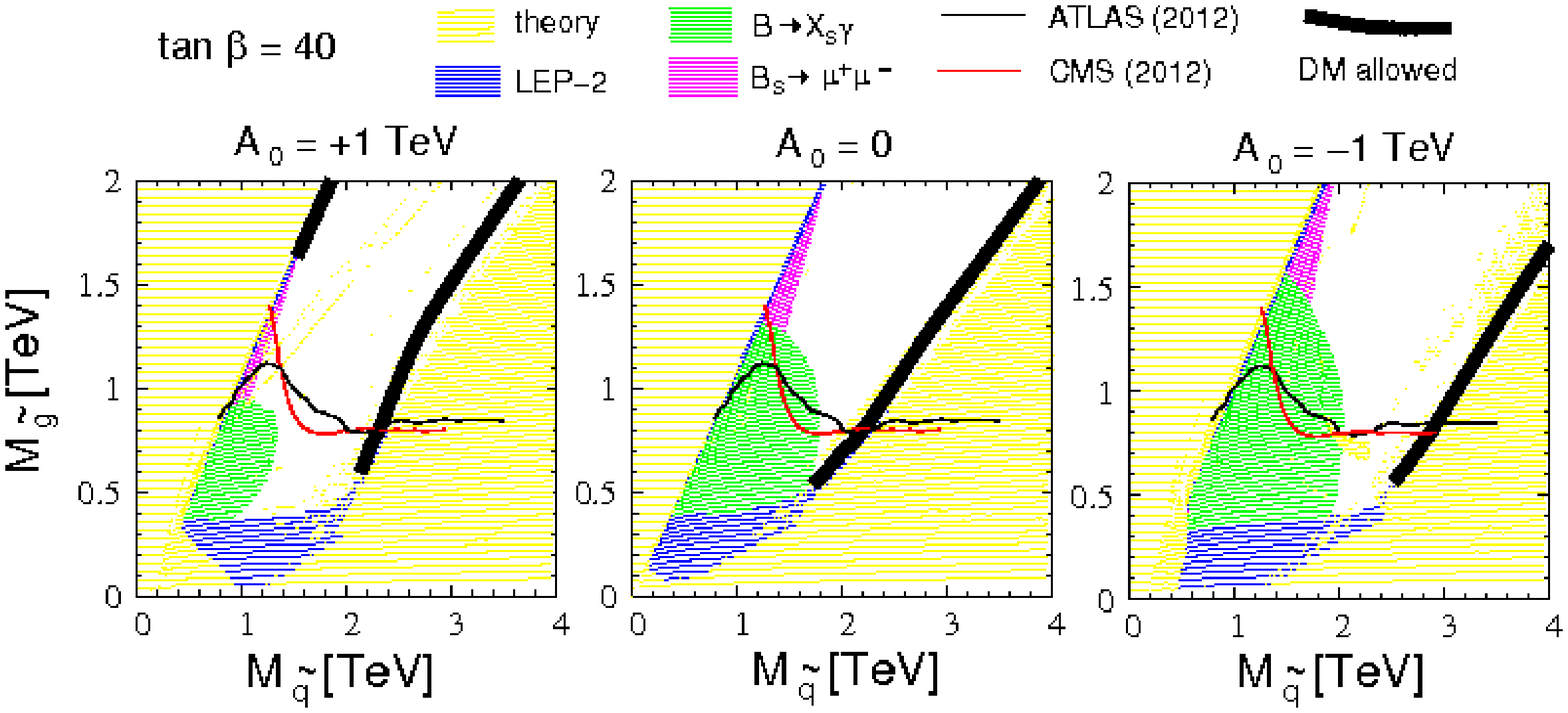} }
\vskip -15pt
\def\baselinestretch{0.8}
\caption{{\footnotesize Illustrating the same constraints as in 
Figure~\ref{fig:mZmH}, but now in the squark and gluino mass plane 
($M_{\widetilde{q}}$--$M_{\widetilde{g}}$ plane). All notations and 
conventions are the same as in Figure~\ref{fig:mZmH}. However, only the 
$\tan\beta = 50$ cases are shown.}}
\def\baselinestretch{1.2}
\label{fig:MqMg}
\end{figure}

Figure~\ref{fig:MqMg} illustrates the constraints in the upper three 
panels of Figure~\ref{fig:mZmH} (i.e. $\tan\beta = 40$), when translated 
into the squark-gluino mass plane. Once again, we use the conventions 
and notations of Figure~\ref{fig:mZmH_10}. The most important feature of 
this graph is the large yellow area ruled out by theory considerations. 
This arises because the squarks (except in the third generation) are 
generally heavier than the gluino in scenarios where the lighter stau is 
heavier than the lightest neutralino. Light gluinos up to a couple of 
hundred GeV appear to be ruled out by the requirement of vacuum 
stability. Higher values of the squark mass cannot break the electroweak 
symmetry unless the gluino is also comparably heavy. There is, then, for 
all values of $A_0$, a funnel-shaped region which is allowed by 
theoretical considerations. Note that theoretically the gluino can be 
substantially lighter for $A_0 \leq 0$ than it is for the $A_0 = +1$~TeV 
case.

As in the previous two figures, we have shown bounds arising from the 
LEP-2 data by blue shading. As we have seen, this arises principally 
from the non-observation of chargino pairs, and this bound on the 
lighter chargino mass translates more-or-less to a constant bound on the 
gluino mass in the ballpark of $300 - 400$~GeV. However, the LHC bounds, 
shown by solid black (ATLAS) and red (CMS) lines as before, are much 
stronger, and they push both the squark and the gluino mass to values 
around a TeV or more. The effect of the low-energy constraints (the 
green and pink-shaded regions) is to marginally constrain some of the 
remaining parameter space. No extra constraint is obtained for $A_0 = 
+1$~TeV, but modest constraints appear for $A_0 \leq 0$, where the 
squark mass is pushed up to at least 1.5~TeV. However, if we consider 
$\tan\beta = 50$ (not shown) most of the allowed region is shut off, and 
for even higher values of $\tan\beta$, nothing is left of it.

\begin{figure}[ht]
\centerline{ \epsfxsize= 6.2 in \epsfysize= 2.8in \epsfbox{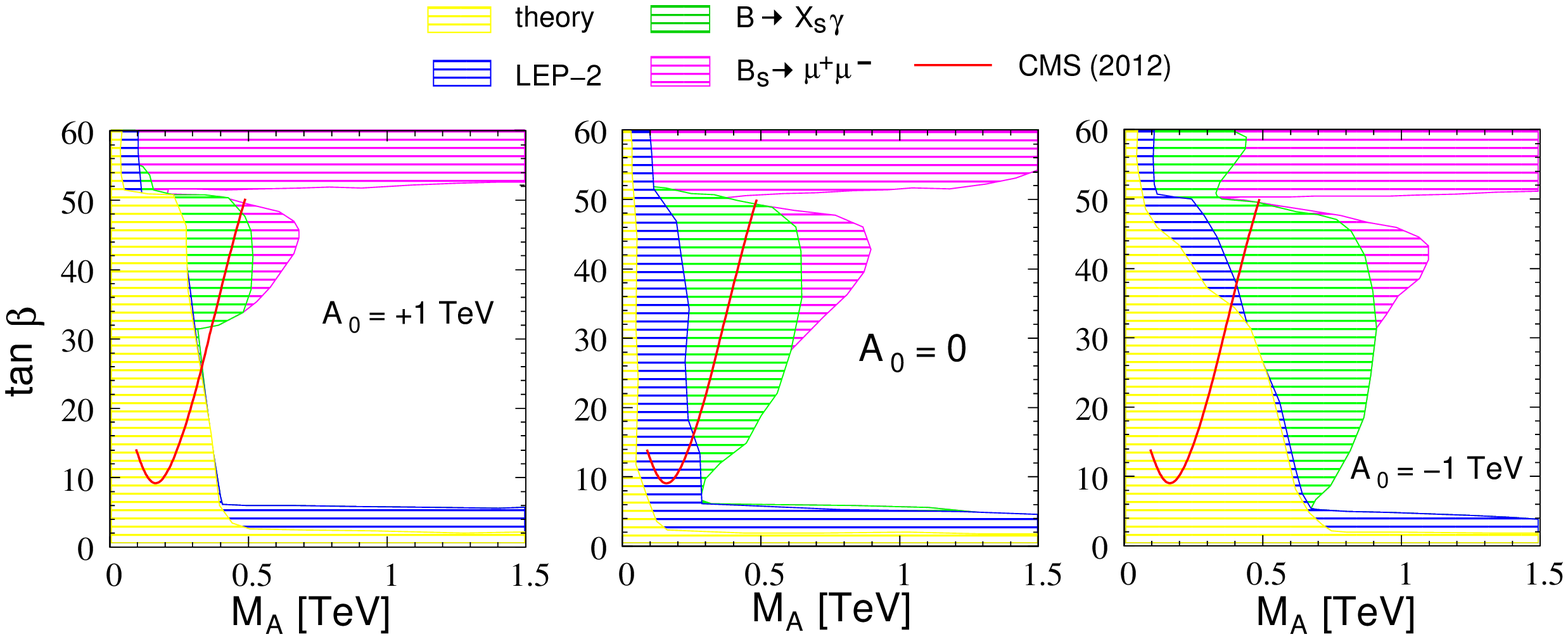} }
\vskip -1pt
\def\baselinestretch{0.8}
\caption{{\footnotesize Illustrating the same constraints as before, but 
now in the $\tan\beta$--$M_A$ plane. All notations and conventions are 
the same as in Figure~\ref{fig:mZmH}, except that $m_0$ and $m_{1/2}$ 
are allowed to float in the same ranges as shown in 
Figure~\ref{fig:mZmH}. In each panel, regions above and to the left of 
the red line are disallowed by direct LHC searches by the CMS 
Collaboration.}}
\def\baselinestretch{1.2}
\label{fig:tBmA}
\end{figure}

The lesson which is learned from the above studies is that while direct 
searches for squarks and gluinos at the LHC produce the same kind of 
constraints for both low and high values of $\tan\beta$ and $A_0$, the 
situation is different for the indirect constraints from low-energy 
measurements, which are generally stronger as $\tan\beta$ increases and 
$A_0$ is driven more strongly negative. To illustrate the full extent of 
this constraint, in Figure~\ref{fig:tBmA}, we have plotted the 
disallowed regions in the plane of $\tan\beta$ and $M_A$, where $M_A$ is 
the mass of the physical pseudoscalar $A^0$.  In this Figure, as in the 
earlier ones, we show three panels for $A_0 = +1, 0$ and -1~TeV 
respectively (from left to right) and set $\mu > 0$ throughout. The 
values of $m_0$ and $m_{1/2}$ are allowed to range from $0 - 4$~TeV and 
$0 - 2$~TeV as before. Of course, for a given value of $A_0$ and 
$\tan\beta$, these cannot vary independently. In fact, as the variation 
of $M_A$ is more directly related to that of $m_0$, one can imagine 
$m_{1/2}$ as the floating variable. Thus, if a point in the 
$\tan\beta$--$M_A$ plane is marked as disallowed, that means that it is 
disallowed for {\it all} values of $m_0$ and $m_{1/2}$ in the box $m0 = 
0 - 4$~TeV and $m_{1/2} = 0 - 2$~TeV.

\vskip -5pt
    
In Figure~\ref{fig:tBmA}, as before, the region shaded yellow indicates 
that it is ruled out by theoretical considerations, or is not accessible 
for the given ranges of $m_0$ and $m_{1/2}$. It is interesting that the 
disallowed region is very small when $A_0 = 0$, but is significantly 
larger when $A_0 \not{\!\!=} \ 0$. This may be traced, as earlier, to a 
larger mixing among the stau gauge eigenstates, leading to a stau LSP. 
The LEP-2 constraints do not change much from panel to panel, which is 
expected, since we have seen that their dependence on $m_0$ is somewhat 
weak. What is of greatest interest in Figure~\ref{fig:tBmA}, however, is 
the regions ruled out by the low-energy constraints. In each case, it is 
clear that for $\tan\beta \geq 50$, the constraint from $B_s \to 
\mu^+\mu^-$ is highly restrictive, effectively pushing the $A^0$ mass to 
the decoupling limit in the Higgs sector. However, this constraint 
becomes ineffective when the value of $\tan\beta$ is lowered, as we have 
already seen.  In this case, however, the constraint from $B \to X_s 
\gamma$ comes into play unless $A_0$ is large, and this has the effect 
of driving the mass of $M_A$ to larger values for intermediate values of 
$\tan\beta$ around 20 -- 45. For low values of $\tan\beta$, the 
low-energy constraints disappear, as we have seen in 
Figure~\ref{fig:mZmH_10}, and we fall back to the LEP-2 constraints. 
Finally, there is a sort of wedge around $\tan\beta = 50$ where $M_A$ as 
low as 500~GeV is allowed by all the constraints. Direct searches for 
the $A^0$ and the charged Higgs bosons $H^\pm$ at the LHC \cite{Asearch} 
lead to the exclusion of points above and to the left of the solid red 
curve --- this is, however, less restrictive than the indirect 
constraints\footnote{It is also relevant to note that these constraints
were derived in the so-called $m_H$-max scenario, which is more restrictive
than the cMSSM.}. If we consider all the diagrams together, we have an 
absolute minimum of around 300~GeV for the $A^0$. This means that the 
charged Higgs boson, which is easier to detect, is of mass around 
310~GeV. In fact, all the heavy scalar states in the cMSSM will now have 
masses of 300~GeV or above, which already makes them difficult to 
detect. In this sector, if not in the sector for SUSY particles, the 
cMSSM is fast approaching the limit where detection at the LHC will no 
longer be possible.
   
\begin{figure}[ht]
\centerline{ \epsfxsize= 3.8in \epsfysize= 2.5in \epsfbox{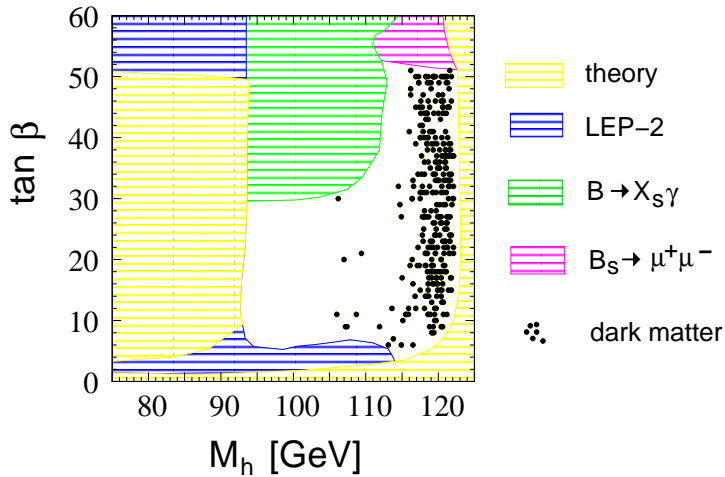} }
\vskip -1pt
\def\baselinestretch{0.8}
\caption{{\footnotesize Illustrating the same constraints as before, but 
now in the $\tan\beta$--$M_h$ plane. All notations and conventions are 
the same as before, except that $A_0$ floats as well as $m_0$ and 
$m_{1/2}$. Note that this region is unaffected by the limits from 
$WW^\ast$ searches at the LHC.}}
\def\baselinestretch{1.2}
\label{fig:tBmh1}
\end{figure}

What about the light scalar state? Obviously, if the heavier scalars 
start approaching their decoupling limit, the lightest scalar $h^0$ will 
also approach its decoupling limit, viz. around $119 - 120$~GeV. The 
exact situation is illustrated in Figure~\ref{fig:tBmh1}, where we plot 
$M_h$ instead of $M_A$, keeping $m_0$ and $m_{1/2}$ floating as before. 
For this plot, we allow $A_0$ also to float over all values from 
$-1$~TeV to $+1$~TeV. As before, a point is marked as disallowed if this 
is valid for all values of $m_0$, $m_{1/2}$ and $A_0$ in the given 
ranges (only two of these are independent, for reasons explained 
before).

\vskip -5pt

In Figure~\ref{fig:tBmh1}, as in the others, the region shaded yellow 
corresponds to the region which is theoretically inaccessible in the 
cMSSM. Of these, the yellow region on the left of the figure arises 
because of the requirement of vacuum stability and convergence of RGE's, 
whereas the yellow region on the right is simply not accessible for the 
parameter range chosen for our study. After all, we must recall that 
$M_h \leq M_Z$ at the tree-level, and hence much of the region shown in 
this plot corresponds to radiative corrections to $M_h$. We shall come 
back to this issue presently. Of the remaining theoretically-accessible 
region in Figure~\ref{fig:tBmh1}, a small portion is ruled out by the 
LEP-2 searches, and comparatively larger regions by the low-energy 
constraints, especially for large values of $\tan\beta$. However, for 
$\tan\beta$ in the range $6 - 30$, these constraints allow for $M_h$ 
anywhere in the region between 93~GeV to about 123~GeV. The lower range 
in $\tan\beta$ is essentially shut off by a combination of theoretical 
constraints and LEP-2 bounds.
 
Much of the above is already well known. The most interesting feature of 
Figure~\ref{fig:tBmh1}, however, is the cluster of black dots, which 
indicates the regions compatible with the dark matter requirement. 
Obviously, these favour a Higgs boson mass in the neighbourhood of $120 
- 122$~GeV, and strongly disfavour the lighter end of the permitted 
region. Interestingly, the favoured region is also close to the 
decoupling regime for the sparticles, and hence, we seem to be looking 
at a strong hint that the sparticles, if found, will turn out to have 
masses well in the ballpark of a few TeV.

Recently, the ATLAS and CMS Collaborations, as well as the Fermilab 
experiments, have reported tantalisingly positive results of their Higgs 
boson searches, which seem to indicate a Higgs boson of mass somewhere 
in the region around 125~GeV. The cMSSM prediction, obviously, lies 
right very close to this. Thus, if this result is confirmed, then, 
broadly speaking, there will be no problem with the health of the cMSSM. 
One can even take pride that a light Higgs boson in this regime has been 
a long-standing prediction of the cMSSM, even in the days when the Higgs 
boson searches allowed for the entire region from $45 - 800$~GeV. 
However, a closer look at Figure~\ref{fig:tBmh1} seems to indicates that 
$M_h = 125$~GeV would be inconsistent with the cMSSM, as we have 
formulated it, since that value lies in what we have called the 
theoretically inaccessible region. However, this is not strictly the 
case, for the upper bound of $M_h \; \lsim \; 123$~GeV which appears 
from the above plot, is really an artefact of our restricting $A_0$ to 
the regime $|A_0| \leq 1$~TeV. Choosing higher values of $A_0$ can 
mitigate the upper limit: for example, if we choose the point
$$
m_0 = 4.0~{\rm TeV} \qquad 
m_{1/2} = 1.0~{\rm TeV} \qquad 
A_0 = -6.0~{\rm TeV} \qquad 
\tan\beta = 25 \qquad 
\mu > 0
$$
we will predict $M_h \simeq 126$~GeV. Obviously, inclusion of a more 
definite constraint from the light Higgs boson could change our present 
study quite drastically \cite{lightHiggs}. For example, if the Higgs boson mass is 
confirmed to lie around 125~GeV, then the parameter space plotted in all 
the five figures in this article will become disallowed, and we will 
only be left with small corners of the parameter space which are 
compatible with the slightly-larger Higgs boson mass. This study cannot 
be carried out at the present juncture, because the parameter space is 
extremely sensitive to the precise value of $M_h$, with large regions 
becoming allowed or disallowed for changes in $M_h$ by as little as 
1~GeV, whereas the error in the theoretical calculations using {\sc 
SuSpect} is around $2 -3$~GeV. The time for such a study will come when 
the higher order corrections to $M_h$ are better understood and more 
precise data are available.

To conclude this section, let us highlight the main results of our 
analysis of the cMSSM parameter space. The main features are

\vspace*{-0.2in} 

\begin{itemize}

\item The cMSSM is still viable in large parts of the parameter space. 
This is especially so for $\tan\beta$ in the range from 10 - 35 and $A_0
> 0$, though there are patches which are allowed even outside these 
ranges. There is no imperative reason, therefore, to write off the cMSSM 
and invoke one or other of its variants.

\item The constraints from low-energy processes such as $B \to X_s 
\gamma$ and $B_s \to \mu^+ \mu^-$ are marginal for large positive $A_0$ 
and only become really effective for large negative $A_0$ and large 
$\tan\beta$. Other low-energy processes yield even weaker constraints. 
The two exceptions are the muon anomalous magnetic moment and the rare 
decay $B^+ \to \tau^+ \nu_\tau$, which together rule out practically all 
of the cMSSM parameter space, except a small region which would be 
accessible to the next set of LHC data analyses.

\item Even when all the constraints are imposed, there are enough 
allowed regions where the cMSSM is compatible with the observed dark 
matter relic density. This will remain true even if the LHC completes 
its run without finding signatures of superparticles. However, if the 
LHC fails to find a light Higgs boson the cMSSM --- as indeed the SM and 
most other supersymmetric models --- will be ruled out.

\item The heavy scalars of the cMSSM are likely to be too heavy to be 
seen at the LHC, at least in the early runs. The light scalar should 
have a mass less than 123~GeV if the SUSY particles are light enough to 
be seen at the LHC. A light scalar $h^0$ with mass around 125~GeV is 
consistent with the cMSSM only in some corners of the parameter space, 
where the superparticles may well turn out to the too heavy to be seen 
at the the LHC. In this case, the cMSSM will still be a possibility, and 
will still constitute and explanation for dark matter, but we will have 
to await a new machine to furnish the experimental proof.

\end{itemize}

\section{Extending the Range of LHC Searches}

\vskip -10pt
In the previous section, we have seen that the parameter space of the 
cMSSM is moderately constrained for large values of $\tan\beta$, but for 
lower values of $\tan\beta$, the overall constraints are somewhat 
modest. The strongest constraints arise in this low $\tan\beta$ regime 
from the direct searches at the LHC. It is important, therefore, to see 
how far these can explore the parameter space. Of course, with increase 
in the energy and integrated luminosity, there will always be 
improvement in the constraints -- unless, indeed, we have a discovery -- 
but further extension of the discovery limits will have to depend on our 
ingenuity in inventing methods to extend the standard searches at the 
LHC into new regions of parameter space.

In this article we take up one such method which was developed by two of 
the authors in Ref.~\cite{dipan} to extend the reach of the jets + MET 
signal which is the major signature of the cMSSM at the LHC. We have 
earlier mentioned that it is convenient to present our results for four 
few benchmark points in the parameter space which we have labelled A, B, 
C and D (see Table~\ref{tab:ABCD}). All of these are permitted by the 
combined set of constraints. The first two, viz. A and B, are compatible 
with neutralino dark matter, while the other two, viz. C and D, are not 
compatible with dark matter, but are nevertheless, allowed by all 
terrestrial experiments\footnote{Except the muon anomalous magnetic 
moment and the decay $B^+ \to \tau^+ \nu_\tau$, which we have not 
implemented in this work for reasons given in the text.}.

The signals at these four benchmark points (BP) will depend crucially on 
the mass spectrum at these points. This is readily generated using the 
package {\sc SuSpect} \cite{suspect}, and we present some of the 
important masses in Table~\ref{tab:spectrum} below.

\begin{table}[htb]
\begin{center}
\begin{tabular}{crrrrrrrrrrrrr}
\hline 
BP & $\tilde{g}$ & $\tilde{q}$  & $\tilde{t}_1$  & $\tilde{t}_2$ & $\tilde{b}_1$ & $\tilde{b}_2$ &
$\tilde{e}_1$ & $\tilde{\tau_{1}}$ & $\tilde{\chi}_{1}^{0}$ & $\tilde{\chi}_{2}^{0}$
&$\tilde{\chi}_{1}^{+} $& $\tilde{\chi}_{2}^{+}$& $h^0$ \\ 
\hline\hline 
A & 1672  & 1530  & 1256  & 1438  & 1395  & 1446 &  634 &  596 & 315 & 590 & 590 & 806 & 116  \\ 
B & 1018  & 2480  & 1533  & 1860  & 1860  & 2113 & 2400 & 2031 & 114 & 156 & 156 & 340 & 115 \\ 
C & 1307  & 1512  & 1080  & 1292  & 1267  & 1353 & 1080 &  873 & 230 & 436 & 436 & 651 & 116 \\ 
D &  938  & 2095  & 1248  & 1557  & 1557  & 1776 & 2003 & 1666 & 148 & 285 & 285 & 530 & 118 \\  \hline
\end{tabular}
\caption{\footnotesize Mass spectrum for the benchmark points A, B, C and D given in 
Table~\ref{tab:ABCD}. All masses are in units of GeV.}
\label{tab:spectrum} 
\end{center}
\end{table}
\vspace*{-0.2in}

The mass spectrum for the BP's labelled A (DM compatible) and C (DM 
non-compatible) broadly resemble each other, while those for B (DM 
compatible) and D (DM non-compatible) are also roughly similar. However, 
for A the gluino is somewhat heavier than the squarks while in the 
others it is the squarks which are heavier than the gluinos. The BP 
labelled A is unique in predicting a light selectron and a light stau 
which is not surprising, since it has been chosen on the edge of the 
(excluded) region where we would have a stau LSP. The LSP 
$\tilde{\chi}_1^0$ varies in mass from the low 114~GeV to the high 
315~GeV --- both being consistent with $\tilde{\chi}_1^0$ as the 
principal component of dark matter. For the BP labelled D, the 
cross-section for producing squarks is clearly negligible and it turns 
out that electroweak production modes are dominant 
\cite{Baer:2012ts,Ghosh:2012mc}). As all the squarks are heavier than 
the gluino, the latter will now undergo three-body decays through 
virtual squarks, but the final states are roughly the same.

The production cross-section, for squarks and gluinos at 8~TeV and 
13~TeV for these four benchmark points are given in Table~\ref{tab:cs} 
below. Both the leading order (LO) and next-to-leading order (NLO) 
results are calculated using the software {\sc PROSPINO} 
\cite{prospino}.

\small
\begin{table}[htb]
\begin{center}
\begin{tabular}{r|cccc}
$\sqrt{s}$ &  A                   & B                      & C                      &  D       \\ \hline\hline
8~TeV         & 1.96  (2.58)    & 5.45   (15.4)     &  3.90  (7.21)     & 11.92  (33.64)  \\
13~TeV       & 44.9  (59.7)    & 114.6 (252.6)   &  97.9  (156.0)   &  226.8 (486.5)  \\  \hline
\end{tabular}
\caption{\footnotesize Cross-section at LO in femtobarns for squarks and 
gluinos at the benchmark points A, B, C and D of Table~\ref{tab:ABCD}. 
Numbers in parentheses indicate NLO cross-sections. }
\label{tab:cs} 
\end{center}
\end{table}
\normalsize
\vskip -15pt

A glance at Table~\ref{tab:cs} tells us that with 10~fb$^{-1}$ of 
integrated luminosity at the LHC running at a centre-of--mass energy 8 
TeV, we may expect to produce anything from a few tens to a few hundreds 
of squark/gluino pairs through strong interactions, and thus, there is 
ample scope to detect the signals for these heavy particles decaying 
through cascades to jets and MET.  However, we must recall that all 
these BP's lie {\it outside} the published reach of the ATLAS and CMS 
experiments, which means that these cross-sections are still compatible 
with the SM background. We shall presently describe a novel set of cuts 
which can suppress the SM background leaving this signal largely intact, 
but, for the moment, let us simply note that the cross-sections are at 
least an order of magnitude larger at 13~TeV, as a result of which these 
points are very likely to be accessible even to the standard search 
techniques when the 13~TeV data become available. This is, of course, 
merely a statement of the obvious, and one could argue that it would be 
more relevant to look for other BP's where the standard searches at 
13~TeV are likely to fail. However, we feel that it would be premature, 
at this stage, to make predictions for the machine reach at 13~TeV 
unless we have a better idea of the possible integrated luminosity. For 
the moment, therefore, we concentrate our efforts on the 8~TeV 
predictions.

Once produced, the squarks and gluinos will undergo cascade decays of 
the typical form
\begin{eqnarray*}
\tilde{g} \ \to \ q  \ ({\rm jet}) \ \ + &\tilde{q}&   \\
             &\hookrightarrow& q  \ ({\rm jet}) + \tilde{\chi}_1^0 \ ({\rm MET})
\end{eqnarray*}
so that a pair of gluinos will lead to a signal with up to four jets and 
MET. Of course, this is only one of many possible configurations, given 
the number and variety of superparticles, but it serves to illustrate 
the origin and importance of the jets plus MET signal. The dominant SM 
backgrounds arise from the production of $t \bar t$ and $W/Z$+jets as 
well as QCD processes --- all having cross-sections some orders of 
magnitude larger than the signal. Eliminating these huge SM backgrounds 
on the way to isolation of the tiny signal is, therefore, a challenging 
task, and requires a very finely-tuned set of cuts, which will 
correspond closely to the signal configurations. We now describe our 
analysis in some detail, mentioning these cuts as and when relevant.

We have used the well-known Monte Carlo event generator {\sc 
Pythia}~\cite{pythia} to simulate the signal events as well as 
backgrounds arising from QCD and $t \bar t$ production. Branching ratios 
for superparticles are calculated using {\sc SUSYHIT} 
\cite{Djouadi:2006bz}. For backgrounds arising from $W/Z$ + jets and 
$t\bar{t}$ + jets we have used {\sc AlpGen}~\cite{alpgen} interfaced 
with {\sc Pythia} for showering adopting MLM~\cite{mlm} matching to 
avoid double counting of jets. We then use {\sc FastJet}\cite{fastjet} 
to obtain jets out of stable particles using the anti-$k_T$ \cite{salam} 
algorithm with cone size $\Delta R = 0.5$. Jets ($J$) are selected with 
a minimum transverse momentum satisfying $p_T^J > 50$~GeV and with 
pseudorapidity satisfying $|\eta_J| < 3$. We use CTEQ6L as PDF from the 
LHAPDF package \cite{cteq}. The missing energy and momentum are 
estimated, as usual, by summing the energy and momentum of all visible 
particles.

In addition to these simple acceptance cuts, we apply cuts on the 
following set of kinematic variables to suppress the large backgrounds 
in the problem, following earlier work by two of the 
authors~\cite{dipan}.

\vspace*{-0.2in}
\begin{enumerate}

\item {\it Transverse events shape variable}: 
The transverse events shape variable $T$ is defined as, 
\begin{equation}
T = {\rm max}_{\vec{n}_T}\frac { \sum_J |\vec q_T^J \cdot \vec n_T| }
{\sum_J q_T^J },
\label{eq:tht}
\end{equation}
where the summation runs over all jets $J$ in the event, $\vec q_T^J$ is 
the transverse component of each jet $J$ and $\vec{n}_T$ is the 
transverse vector which maximizes this ratio.

As discussed in Ref.~\cite{dipan}, for events with low jet multiplicity, 
such as those belonging to the SM backgrounds, we expect to have $T \sim 
1$, whereas for events with high jet multiplicity, such as those 
belonging to the signal, we expect to have a significant deviation $T < 
1$. Therefore, an upper cut on the $T$ variable eliminates a good 
fraction of the background events without paying much in loss of signal 
events. In our analysis, we get optimum results by requiring $T \leq 
0.9$.

\item {\it Ratio of sums of jet $p_T$}: 
We define the ratio
\begin{equation}
R_{T}\left(n_J^{\rm min}\right) = \frac{1}{H_T} \sum_{J = 1}^{n_J^{\rm 
min}} p_T^J
\label{eq:RT}
\end{equation}
where $H_T = \sum_{J = 1}^{n_J} p_T^J$, with $n_J^{\rm min}$ being the 
lowest number of jets required to trigger the events and $n_J$ being the 
actual number of jets in the event.

Obviously, events where $n_J \sim n_J^{\rm min}$ can be expected to 
yield $R_T \simeq 1$ whereas $n_J \gg n_J^{\rm min}$ will result in a 
significant deviation $R_T < 1$ as was observed in Ref.~\cite{dipan}. 
Hence, as before, a cut on the maximum $R_T(n_J^{\rm min})$ will 
suppress the SM backgrounds quite effectively with only a modest loss in 
signal events. We optimise this as $R_T(4) \leq 0.85$.

\item {\it Transverse invariant mass between the two leading jets}: We 
define this as
\begin{equation}
M_T^{12} = \left[ 2p_T^{J_1} p_T^{J_2} \left(1 - \cos\phi_{12}\right)\right]^{1/2} 
\end{equation}
where $J_1$ and $J_2$ are the leading (highest $p_T^J$) jets and 
$\phi_{12}$ is the azimuthal angle between them.

This variable is defined primarily to eliminate backgrounds from $t\bar 
t$ events where one would expect the two leading jets to be widely 
separated, resulting in small values of $M_T^{12}$. However, in the 
signal events, the parent particles are comparatively heavier and hence 
the jets are more evenly distributed. Therefore, demanding larger values 
of $M_T^{12}$ would remove a sizeable fraction of the $t\bar{t}$ 
backgrounds with only marginal effects on the signal events. In our 
numerical analysis, we have set this criterion as $M_T^{12} \geq 
450$~GeV.

\item {\it Missing transverse momentum}: As in all searches for 
supersymmetry, a cut $p{\!\!\!/}_T \ge $ 200~GeV is required to get rid 
of the remaining background events.

\end{enumerate}

\begin{table}[htb]
\begin{center}
\begin{tabular}{r|rrrr|rrrr|r}
\hline
Cut & A & B & C & D & 
$t\bar{t}$ + jets & QCD & $W^\pm$ + jets & $Z^0$ + jets & Total SM \\
\hline\hline
Produced & 19 & 55 & 39 & 119 & 1\,384\,500 & 3\,370\,100 & 6\,180\,000 &  
1\,422\,000 & 22\,356\,600 \\
$T \leq 0.9$ & 10 & 45 & 30 & 101 & 593\,978 & 746\,666  &
2\,843\,521 & 768\,130 & 5\,952\,295 \\
$R_T(4) \leq 0.85$  &  3  & 24 & 17 & 62 & 29\,928 & 8\,496 & 1\,558 & 148 &     110\,129 \\
$M_T^{12} \geq 450$~GeV & $< 1$ & 13 & 11 & 26 & 907 & 1263 & 164 & 11 & 32\,344 \\
$p{\!\!\!/}_T \ge $ 200~GeV &  $< 1$  &  7  &  8 &  13 & 6  & $< 1$ & $< 1$ & $< 1$ 
& 6 \\
\hline
\end{tabular}
\caption{\footnotesize Cut flow table for the four BP's A--D and the 
principal backgrounds. The cross-sections are generated at LO using {\sc 
PROSPINO} for $\sqrt{s} = 8$~TeV and the number of events in each entry 
is for an integrated luminosity of 10~fb$^{-1}$. }
\label{tab:cutflow} 
\end{center}
\end{table}
\vskip -15pt

The effect of these cuts on the signal at the four BP's and on the 
principal backgrounds is illustrated in Table~\ref{tab:cutflow}, where 
we have assumed $\sqrt{s} = 8$~TeV and an integrated luminosity of 
10~fb$^{-1}$. The first row shows the number expected using the LO 
cross-section using {\sc PROSPINO} \cite{prospino} for the signal 
process. The effect of applying the four special cuts is then apparent 
from the successive rows. It is obvious that, taken together, these cuts 
are extremely effective, reducing the SM backgrounds by seven orders of 
magnitude. In fact, most of the SM backgrounds are removed completely, 
with the small irreducible portion coming entirely from the $t\bar{t}$ 
production.

If we consider the numbers presented in Table~\ref{tab:cutflow}, it is 
obvious that the cut on $T$ is most effective in suppressing the massive 
QCD background, while the cuts on $R_T(4)$ and $M_T^{12}$ suppress all 
the backgrounds, but those from vector boson production more than the 
others. The final cut on $p{\!\!\!/}_T$ removes all the backgrounds 
except a small irreducible part from $t\bar{t}$, always the principal 
bugbear for most new physics studies. However, the combined effect of 
all the cuts is very impressive, for it reduces a background of a 
hundred million events to just sixty events, whereas a sizeable fraction 
of the the few tens of signal events remains.
 
The last line of Table~\ref{tab:cutflow} tells us that it is quite 
hopeless, at $\sqrt{s} = 8$~TeV, to search for the BP labelled A. In 
fact, it has been observed \cite{dipan} that this search technique works 
better as we go to larger values of $m_0$, and we observe that A has the 
lowest $m_0 = 0.4$~TeV. For the other BP's, however, if we neglect 
systematic effects, we should expect to see a signal at a confidence 
level more than 95\% in all cases. Thus, using the present technique, we 
can hope for ($a$) either a discovery of the cMSSM in the jets + MET 
channel during the current year, or at least ($b$) a significant 
improvement in the LHC exclusion regions before the machine closes down 
for the upgrade to 13~TeV. This latter would narrow down the cMSSM 
search region further, but would not invalidate the cMSSM in any way.
      
\section{Summary}

\vskip -10pt
In this article we have critically studied the cMSSM, stressing its 
strong and weak points, except for its flavour aspects --- which call 
for a different kind of study. In the introductory section we have 
reiterated that the cMSSM is arguably still the best option where 
extensions of the SM are concerned, and therefore one should have convincing 
empirical proof of its demise before other supersymmetric models can be 
taken seriously. The cMSSM possesses not just a whole array of pleasing 
naturalness features, but it also provides an elegant explanation of 
electroweak symmetry-breaking, which is one of the many artificial 
contrivances in the SM. The other feature which makes the CMSSM so 
attractive is that it provides an excellent candidate for a particulate 
nature of non-baryonic dark matter, which we now know for certain to 
form a major component of the Universe.

We have then investigated the implications of existing constraints on 
the cMSSM. These turn out to be of four kinds, viz. ($i$) {\sl theory 
constraints} which arise from the cMSSM becoming inconsistent with 
electroweak symmetry-breaking, ($ii$) {\sl indirect constraints} which 
require consistency with low-energy data, especially those coming from 
B-physics experiments, ($iii$) {\sl direct constraints} which arise from 
searches at colliders, of which the LHC searches explore the maximum 
region of parameter space, and ($iv$) {\sl dark matter constraints} 
which arise from the requirement that the observed dark matter is 
composed of neutralinos. Interestingly, each of the constraints comes 
with a caveat. The theory constraints affect most those parts of the 
parameter space where observable signals at colliders are expended; they 
can be easily evaded at the expense of falsifiability. The indirect 
constraints arise from virtual superparticle contributions, which can 
always be mimicked/cancelled by other new physics contributions. 
Collider searches and a stable neutralino require the conservation of 
$R$-parity, which is not demanded by the supersymmetry, but which is put 
into the cMSSM as a phenomenological nicety. Over and above these, there 
obviously will be no constraint if there is some other explanation of 
the dark matter effects.

Despite all these caveats, each of which is equivalent to some level of 
wishful thinking, we have taken the cMSSM with the whole package of 
assumptions which go with these caveats, and explored the exact status 
of its rather limited parameter space when these constraints are applied 
on it, both individually and in unison. Quite surprisingly, it turns out 
that the parameter space is only peripherally affected by all these 
constraints and prejudices. It is only the lower and upper ends of the 
range in $\tan\beta$ which are absolutely ruled out: the lower end by 
direct collider searches at LEP-2, and the upper end by a combination of 
constraints from the low-energy processes $B \to X_s \gamma$ and $B_s 
\to \mu^+ \mu^-$, with the latter making a bigger impact at high values 
of $\tan\beta$. In the intermediate region, say $\tan\beta \sim 5 - 35$, 
the only serious constraints come from theory and the direct searches. 
These leave large areas of the parameter space unexplored, but it is now 
more-or-less clear that the first two generation squarks and the gluinos 
of the cMSSM must lie close to or above a TeV in mass. Somewhat higher 
masses represent no problem for the cMSSM, as a model, but will require 
more running of the LHC before they can be probed. So far as the 
superparticle searches are concerned, the results of the early LHC runs 
can be concisely expressed as ($i$) a proof that a cMSSM discovery was 
not `around the corner' when the LHC was turned on, and ($ii$) that high 
values of $\tan\beta \gsim 50$ are disfavoured.

One of the most interesting features of the constraints on the cMSSM is that
so long as some region of the parameter space is permitted, there seems to be
a set of points which are compatible with the neutralino explanation of dark 
matter. In fact, our plots show that even if the LHC completes its run without
finding signals for sparticles, there will still exist portions of the parameter 
space which allow for neutralino dark matter. This is not an issue, therefore,
that can be settled by the direct searches.  

Perhaps the strongest constraints on the cMSSM will eventually come from 
the most tenuous of the existing empirical information, viz. the hints 
for a 125~GeV Higgs boson. For a SM Higgs boson, we have compelling 
evidence to believe that it cannot lie elsewhere than in the range $115 
- 127$~GeV. Moreover, there are some excess events over background in 
several channels for a Higgs boson of mass around 125~GeV. We must 
remember, however, that the light Higgs boson of the cMSSM is not quite 
the SM Higgs boson, since there are differences in their couplings to 
other fields, such as heavy gauge bosons and fermions. Thus the SM Higgs 
search results cannot be naively taken over into supersymmetry.  We have 
shown that a combination of the existing constraints (and prejudices) 
indicates a light Higgs boson in the range of around $93 - 123$~GeV with 
the upper part of the range being favoured by dark matter. A light Higgs 
boson lying in the lower part of this region (say, around 100~GeV), 
would, therefore, be compatible with all terrestrial experiments, but 
would require serious re-thinking about the composition of non-baryonic 
dark metter. On the other hand, a light Higgs boson with mass somewhat 
higher than 123~GeV would require us to invoke portions of the parameter 
space which have been left unexplored in the present work, such as very 
high values of $A_0$. Here, every GeV counts -- an increase in the light 
Higgs boson mass by a single GeV would rule out large swaths of the 
parameter space and vice versa. Thus, it is important to know the exact 
experimental constraint on the light Higgs boson of the cMSSM which 
arises from the LHC data. Till such results become available, we can but 
speculate.

Looking ahead, it seems unlikely that in the near future, low energy 
constraints from $B$-physics will improve so dramatically as to make 
more than incremental changes to the disallowed parameter space as shown 
in this work. The strongest constraints on the cMSSM parameter space, 
will, therefore, now come from the LHC direct search data -- at 8~TeV, 
as well as the upgraded run at 13~TeV. It is important, therefore to see 
just how much of the cMSSM parameter space can be accessed by studying 
these data. In this context, a search study using event-shape variables 
has been shown to have a longer reach than other conventional searches. 
We have, therefore, showcased this by making a detailed study at some 
chosen benchmark points in the parameter space, which lie in the region 
allowed by the present data. We show that these could be accessed by our 
novel study method, and hence, the eventual range of accessible 
parameter space at the upgraded energy may also be expected to grow 
accordingly.

All in all, if we consider all the serious empirical evidence available 
at the present juncture, the cMSSM is still in pretty good health, even 
though some extremities of the allowed parameter space have been lopped 
off. However, when more results of the current searches for the Higgs 
boson become available, it may be tested more stringently than it ever 
has been in the past. Failure to find a light Higgs boson altogether 
will certainly spell doom for the cMSSM, as indeed, it will for the SM 
and the entire picture of electroweak symmetry-breaking which has been 
built up over the past half-century. A Higgs boson discovery, while 
vindicating this picture, will be only the start of the real search for 
supersymmetry, since it will not only make supersymmetry a theoretical 
necessity, but also provide crucial information on the corner of 
parameter space in which to search for the superparticles. Anticipating 
such a discovery, we have explored the potential of a novel 
background-elimination technique to explore regions in the parameter 
space where conventional searches cannot penetrate. Whether these will 
be required or not, however, is a matter which only the future can 
decide. On this our case rests.

\begin{quotation}
\def\baselinestretch{1.0}
\noindent\small {\it Acknowledgements}: DG thanks the Indian Association 
for Cultivation of Sciences, Kolkata, for hospitality while part of this 
work was being done.
\def\baselinestretch{1.2}
\end{quotation}
\normalsize


\end{document}